\documentclass[11pt]{article}
\usepackage{authblk}

\usepackage{graphicx}
\usepackage{subfig}
\usepackage{amsmath}
\usepackage{amsfonts}
\usepackage{amssymb}
\usepackage{enumerate}
\usepackage{lscape}
\usepackage{longtable}
\usepackage{color}
\usepackage{url}
\usepackage{algorithm}
\usepackage{algpseudocode}
\usepackage{setspace}
\usepackage{caption}

\usepackage{array}

\usepackage[margin=1in]{geometry}

\newcolumntype{M}[1]{>{\centering\arraybackslash}m{#1}}
\newcolumntype{N}{@{}m{0pt}@{}}

\newtheorem{theorem}{Theorem}

\newtheorem{definition}[theorem]{Definition}
\newtheorem{lemma}[theorem]{Lemma}
\newtheorem{problem}[theorem]{Problem}
\newtheorem{proposition}[theorem]{Proposition}

\def\EE{\mathbb{E}}

\renewcommand{\iota}{{l}}

\newcommand{\argmin}{\operatorname{argmin}}

\newcommand{\PP}{\mathrm{Pr}}
\newcommand{\RR}{\mathbb{R}}

\title{Fast Cross-Polytope Locality-Sensitive Hashing}

\author{Christopher Kennedy\thanks{Email: \texttt{ckennedy@math.utexas.edu}. C. Kennedy was supported in part by R. Ward's NSF CAREER grant and an ASOFR Young Investigator Award} }

\author{Rachel Ward\thanks{Email: \texttt{rward@math.utexas.edu}.  R. Ward was supported in part by an NSF CAREER grant and an ASOFR Young Investigator Award}}

\affil{Department of Mathematics, University of Texas at Austin}

\begin{document}
\maketitle
\onehalfspacing

\abstract
We provide a variant of cross-polytope locality sensitive hashing with respect to angular distance which is provably optimal in asymptotic sensitivity and enjoys $\mathcal{O}(d \ln d )$ hash computation time.  Building on a recent result in \cite{pracoptlsh}, we show that optimal asymptotic sensitivity for cross-polytope LSH is retained even when the dense Gaussian matrix is replaced by a fast Johnson-Lindenstrauss transform followed by discrete pseudo-rotation, reducing the hash computation time from $\mathcal{O}(d^2)$ to $\mathcal{O}(d \ln d )$.  Moreover, our scheme achieves the optimal \emph{rate of convergence} for sensitivity.  By incorporating a low-randomness Johnson-Lindenstrauss transform, our scheme can be modified to require only $\mathcal{O}(\ln^9(d))$ random bits.

\newpage

\section{Introduction}

The nearest neighbor search problem is an essential algorithmic component to a wide variety of applications including data compression, information retrieval, image storage, computer vision, and pattern recognition.  \textbf{Nearest neighbor search (NN)} can be stated as follows: given a metric space $(X, \mathcal{D})$ and a set of points $P = \{x_1,...,x_n\} \subset X$, for a query point $x \in P$ find $y = \argmin_{x_i \in P \setminus \{x\}} \mathcal{D}(x_i,x)$.  In high dimensions, it is known that existing algorithms have poor performance (see \cite{weber1998quantitative}); that is, for a query point $x \in P$, any algorithm for NN must essentially compute the distances between $x$ and each point in $P \setminus \{x\}$.

In order to improve on linear search, one may relax the problem to that of \emph{approximate} nearest neighbors search.  Precisely, the \textbf{$(R,c)$ near neighbor problem ($(R,c)$-NN)} as introduced in \cite{ApproxNN} is as follows: given a query point $x \in P$ and the assurance of a point $y^{\prime} \in P$ such that $\mathcal{D}(y^{\prime},x) <R$, find $y \in P$ such that $\mathcal{D}(y,x) < cR$.  In contrast to exact nearest neighbors search, the approximate nearest neighbor search problem can be solved in \emph{sublinear} query time, and this is achieved using \textbf{locality sensitive hashing} (LSH).   The idea in LSH  is to specify a function from the domain $X$ to a discrete set of hash values -- a so-called \emph{hash function} -- which sends closer points to the same \emph{hash value} with higher probability than points which are far apart.  Then, for a set of points $P = \{x_1,...,x_n\} \subset X$ and a query point $x \in P,$ search within its corresponding hash bucket for a nearest neighbor.  

From here on out, we fix the space $X = S^{d-1}$ endowed with the euclidean metric.  We begin by recalling the standard notion of sensitivity for a hash family; intuitively, a hash family with higher sensitivity is much more likely to hash points that are close to the same hash value, and thus be a better candidate for locality sensitive hashing.

\begin{definition}
For $r_1 \leq r_2$ and $p_2 \leq p_1$, a hash family $\mathcal{H}$ is \textbf{$(r_1,r_2,p_1,p_2)$-sensitive} if for all $x,y \in S^{d-1}$,
\begin{itemize}
\item If $\|x-y\|_2 \leq r_1$, then $\PP_{\mathcal{H}}[h(x) = h(y)] \geq p_1$.
\item If $\|x-y\|_2 \geq r_2$, then $\PP_{\mathcal{H}}[h(x) = h(y)] \leq p_2$.
\end{itemize}
\end{definition}
We primarily care about the case where $r_1 = R$, $r_2 = cR$, and to quantify sensitivity of a certain scheme, we study the parameter
\begin{equation}\label{def_rho}
\rho  = \frac{\ln(p_1^{-1})}{\ln(p_2^{-1})}.
\end{equation}
The key result linking the sensitivity of a hash family to its performance for $(R,c)-NN$ search is the following:\footnote{In particular, the algorithm stores $L$ hash tables from the family $\mathcal{G}$, where each $g\sim \mathcal{G}$ is given by $g(x) = (h_1(x),...,h_k(x))$, and $h_i \sim \mathcal{H}, i=1...k$.  Then, given a query point $x \in S^{d-1}$, the algorithm looks for collisions in the buckets $g_1(x),...,g_L(x)$.  The choice of parameters $k = n^{\rho}$, $L = \ln_{1/p_1} n$ ensure that the algorithm solves $(R,c)-NN$ with constant probability.}

\begin{proposition}[Theorem 5 in \cite{ApproxNN}] Given an $(R,cR,p_1,p_2)$-sensitive hash family $\mathcal{H}$, there exists a data structure that solves $(R,c)-NN$ with constant probability using $\mathcal{O}(dn + n^{1+\rho})$ space, $\mathcal{O}(n^{\rho})$ query time, and $\mathcal{O}(n^{\rho} \ln_{1/p_1} n)$ evaluations of hash functions from $\mathcal{H}$.
\end{proposition}

Since the parameter $\rho$ quantifies the performance of a given LSH algorithm for $(R,c)-NN$, it is of interest to make this parameter as small as possible.  It was shown in \cite{optlblsh} that $\rho = \frac{1}{c^2}$ is asymptotically (in $d$) optimal for the case of unit sphere with the euclidean metric.  Spherical LSH (\cite{BeyondLSH}, \cite{andoni2015optimal}) was shown to achieve this optimal sensitivity; however, the corresponding hash functions in spherical LSH are not practical to compute.  Subsequently, Andoni, Indyk, Laarhoven, and Razenshteyn \cite{pracoptlsh}  showed the existence of an LSH scheme with optimally sensitive hash functions which are practical to implement; namely, the \emph{cross-polytope} LSH scheme which has been previously proposed in \cite{terasawa2007spherical} (see also \cite{lowerbound}, \cite{optlblsh}, \cite{lblsh}).  Given a matrix ${\cal G} \in \RR^{d\times d}$ with i.i.d. $\mathcal{N}(0,1)$ entries, the cross polytope hash of a point $x \in S^{d-1}$ is defined as 
\begin{equation}\label{cphash_eq}
h(x) = \underset{u=\{\pm e_i\}}{\argmin} \left\| \frac{{\cal G}x}{\|{\cal G}x\|_2} - u \right\|_2,
\end{equation}
where $\{e_i\}_{i=1}^d$ is the standard basis for $\RR^d$.  The paper \cite{pracoptlsh} provided the following collision probability for cross-polytope LSH. 
\begin{proposition}[Theorem 1 in \cite{pracoptlsh}] \label{cpcollprob_theorem}  Suppose $x,y \in S^{d-1}$ are such that $\|x-y\|_2 = R$, with $0< R < 2$, and $\mathcal{H}$ is the hash family defined in (\ref{cphash_eq}). Then,
\begin{equation} \label{cphashcollprob_eq}
\ln \left( \frac{1}{\PP_{\mathcal{H}} [ h(x) = h(y)]} \right) = \frac{R^2}{4-R^2} \ln d + \mathcal{O}_R( \ln (\ln d)).
\end{equation}
Consequently,
\[
\rho = \frac{1}{c^2} \frac{4-c^2 R^2}{4- R^2} + o(1),
\]
\end{proposition}
where here and in the sequel, $o(1)$ means a parameter that goes to 0 as $d\to \infty$.  This implies that the above scheme is asymptotically optimal with respect to $\rho$.\footnote{In fact, the coefficient $\frac{4 - c^2 R^2}{4 - R^2} < 1$ for every choice of $c>1$ and $0<R<2$, but this does not break the lower bound given in \cite{optlblsh} since the lower bound $\rho = \frac{1}{c^2}$ only holds for a particular sequence $R = R(d)$.  For cross-polytope LSH and the schemes proposed here, any sequence $R(d) \to 0$ suffices.}  Still, this scheme is limited in efficiency by the $\mathcal{O}(d^2)$ computation required to compute a dense matrix-vector multiplication in \eqref{cphash_eq}.   To reduce this computation, \cite{pracoptlsh} proposed to to use a pseudo-random rotation in place of a dense Gaussian matrix, namely,
\begin{equation}
\label{fast_prior}
h(x) = \argmin_{u = \{ \pm e_i\} } \left\| H D_b H D_{b^{\prime}} H D_{b^{\prime \prime}} x - u \right\|_2,
\end{equation}
where $H \in \RR^{d\times d}$ is a Hadamard matrix and $D_b, D_{b'}, D_{b''} \in \RR^{d \times d}$ are independent diagonal matrices with i.i.d. Rademacher entries on the diagonal.  This scheme has the advantage of computing hash functions in time $\mathcal{O}(d \ln d),$ and was shown in \cite{pracoptlsh} to \emph{empirically} exhibit similar collision probabilities to cross-polytope LSH, but provable guarantees on the asymptotic sensitivity of this fast variant of the standard cross-polytope LSH remain open.

\subsection{Our Contributions}

\subsubsection{Fast cross-polytope LSH with optimal asymptotic sensitivity}
While we do not prove theoretical guarantees regarding the asymptotic sensitivity of the particular fast variant \eqref{fast_prior}, we construct a different variant of the standard cross-polytope LSH (defined below in \eqref{fastcphash_eq}) which also enjoys $\mathcal{O}(d \ln d )$ matrix-vector multiplication, and for which we are able to prove optimal asymptotic sensitivity $\rho = \frac{1}{c^2}$:
\begin{equation} \label{fastcphash_eq}
h_F(x) =  \underset{u=\{\pm e_i\}}{\text{argmin}} \left\| \frac{{\cal G} (H_S D_b x)}{ \|{\cal G} (H_S D_b x)\|_2} - u \right\|_2;
\end{equation}
Here, $D_b : \RR^d \to \RR^d$ is a diagonal matrix with i.i.d. Rademacher entries on the diagonal, $H_S \in \RR^{m \times d}$ is a partial Hadamard matrix restricted to a random subset $S \subset [d]$ of $|S| = m = \mathcal{O}(\log(d))$ rows, and $\mathcal{G} : \RR^m \to \RR^{d^{\prime}}$ is a Gaussian matrix that lifts and rotates in dimension $d^{\prime}$ in the range $m \leq d' \leq d$.  There is nothing special about lifting to dimension $d$, and indeed one could lift to dimension $d^{\prime} > d$, but if $d^{\prime}$ grows faster than $d$, the hash computation no longer takes time $\mathcal{O}(d \ln d)$.

The embedding $H_S D_b x$ acts as a Johnson-Lindenstrauss (JL) transform\footnote{Formally, given a finite metric space $(X,\|\cdot\|) \subset \RR^d$, a JL transform is a linear map $\Phi: \RR^d \to \RR^m$ such that for all $x \in X$, $(1-\delta)\| x \|^2 \leq \|\Phi x\|^2 \leq (1+\delta) \|x\|^2 $, with $m \ll d$ close to the optimal scaling $m = C \delta^{-2} \ln(|X|)$ \cite{johnson1984extensions, alon03, larsen2014johnson}.}, and embeds the points in  dimension $m \approx \ln d$. 

  It is straightforward that the hash computation $x \to h_F(x)$ takes $\mathcal{O}(d^{\prime} m)$ time from the Gaussian matrix multiplication and  $\mathcal{O}(d \ln d)$ time from the JL transform.  We will show that optimal asymptotic sensitivity is still achieved without lifting,  $d^{\prime} = m,$ but we observe both empirically and theoretically that the \emph{rate of convergence} to the asymptotic sensitivity improves by lifting to higher dimension; taking $d^{\prime}$ closer to $d$ results in empirically closer results to the standard cross-polytope scheme (see section \ref{sect_numerexp} for more details).  Moreover, our scheme achieves the lower bound given by Theorem 2 in \cite{pracoptlsh} for the fastest rate of convergence among all hash families which has to $d^{\prime}$ values.

\subsubsection{Fast cross-polytope LSH with optimal asymptotic sensitivity and few random bits}

Aiming to construct a hash family with similar guarantees which also uses as little randomness as possible, we also consider a discretized version of the fast hashing scheme \eqref{fastcphash_eq} in which the Gaussian matrix $\mathcal{G} \in \RR^{d^{\prime} \times m}$ is replaced by a matrix $\widehat{\cal{G}} \in \RR^{d^{\prime} \times m}$ whose entries are i.i.d. discrete approximations of a Gaussian; in place of the ``standard" fast JL transform $H_S D_b$,  we consider $Z \in \RR^{d \times m}$ a low-randomness JL transform that we will clarify later.  Then, the discrete fast hashing scheme we consider is

\begin{equation} \label{fastdiscretehash_eq}
h_D(x) = \underset{u=\{\pm e_i\}}{\argmin} \left\| \frac{\widehat{{\cal G}} (Z x) }{\| \widehat{{\cal G}} (Z x) \|_2} - u\right\|_2.
\end{equation}

Also for this scheme, the hash computation $x \to h(x)$ takes $\mathcal{O}(d^{\prime} m)$ time from the Gaussian matrix multiplication and  $\mathcal{O}(d \ln d)$ time from the JL transform.  
Our scheme has several advantages, due to the fact that the choice of $d^{\prime}$ in the range $d \leq d' \leq m$ is flexible:
To summarize our main contributions, we prove for both the fast cross-polytope LSH and the fast discrete cross-polytope LSH,
\begin{itemize}
    
    \item For each $d^{\prime}$ in the range $m \leq d^{\prime} \leq d$, this scheme achieves the asymptotically optimal $\rho$.  Moreover, for $d^{\prime} = d$, the rate of convergence to this $\rho$ is optimal over all hash families with $d$ hash values.
    
    \item With the choice $d^{\prime} = d$, the scheme computes hashes in time $\mathcal{O}(d \ln d)$ and performs well empirically compared to the standard cross-polytope with dense Gaussian matrix (see section \ref{sect_numerexp}).
    
    \item With the choice $d^{\prime} = m$, and by discretizing the Gaussian matrix, we arrive at a scheme that has only $\mathcal{O}(\ln^9(d))$ bits of randomness and still has optimal asymptotic sensitivity.
    
\end{itemize}

Table \ref{table_hashfn} contains the construction of the original cross-polytope LSH scheme, our fast cross-polytope scheme, as well as the discretized version.

\begin{table}[h!]
\centering
\caption{{ \small Various LSH Families and corresponding Hash Functions. }}

\begin{tabular}{ || M {4cm} | M {9cm} | N||}
    \hline
    LSH Family  & Hash Function &\\[5pt]
    \hline
    Cross-Polytope LSH & {\large $h(x) = \underset{u=\{\pm e_i\}}{\argmin} \left\| \frac{{\cal G}x}{ \| {\cal G}x \|_2} - u \right\|_2$, \hspace{1mm} ${\cal G} \in \RR^{d\times d}$} &\\[30pt]
    \hline
   {\bf  Fast Cross-Polytope LSH } & {\large $h_F(x) = \underset{u=\{\pm e_i\}}{\argmin} \left\| \frac{{\cal G} (H_S D_b x)}{ \|{\cal G}(H_s D_b x)\|_2} - u \right\|_2$, \hspace{1mm} ${\cal G} \in \RR^{d^{\prime} \times m}$} &\\[30pt]
    \hline
   {\bf  Fast Discrete Cross-Polytope LSH} & {\large $h_D(x) = \underset{u=\{\pm e_i\}}{\argmin} \left\| \frac{\widehat{{\cal G}} (Z x) }{\| \widehat{{\cal G}} (Z x) \|_2} - u\right\|_2$, \hspace{1mm} ${\cal \widehat{G}} \in \RR^{d^{\prime} \times m}$} &\\[30pt]
    \hline 
\end{tabular}
\label{table_hashfn}
\end{table}

\bigskip

\subsection{Related work} Many of our results hinge on the careful analysis of collision probabilities for the cross-polytope LSH scheme given in \cite{pracoptlsh}.  Additionally, various ways to reduce the runtime of cross-polytope LSH, specifically using fast, structured projection matrices, are mentioned in \cite{latticesieve}. They also define a generalization of cross-polytope lsh that first projects to a low dimensional subspace, but they never consider lifting back up to a high dimensional subspace again.  Johnson-Lindenstrauss transforms have previously been used in many approximate nearest neighbors algorithms, (see \cite{ApproxNN}, \cite{PracApproxNN}, \cite{FastJLT}, \cite{RandANN},  \cite{BeyondLSH}, and \cite{dasgupta2011fast}, to name a few), primarily as a preprocessing step to speed up computations that have some dependence on the dimension.  LSH with p-stable distributions, as introduced in \cite{indyk2004locality}, uses a random projection onto a single dimension, which is later generalized in \cite{nearopthash} to random projection onto $o(\ln d)$ dimensions, with the latter having optimal exponent $\rho = \frac{1}{c^2} + \mathcal{O}(\ln(\ln d)/ \ln^{1/3} d)$.  We make a note that our scheme uses dimension reduction slightly differently, as an intermediate step before lifting the vectors back up to a different dimension.

Similar dimension reduction techniques have been used in \cite{compressedhash}, where the data is sparsified and then a random projection matrix is applied.  The authors exploit the fact that the random projection matrix will have the restricted isometry property, which preserves pairwise distances between any two sparse vectors.  This result is notable in that the reduced dimension has no dependence on $n$, the number of points.  See section \ref{sec_OpenProblems} for more discussion.

\section{Notation}
We now establish notation that will be used in the remainder.  $\mathcal{O}_R(f(d))$ is to mean $\mathcal{O}_R(f(d)) = \mathcal{O}(f(d) g(R))$ for some finite valued function $g : (0,2) \to \RR$.  The expression $o(1)$ is a quantity such that $\lim_{d\to \infty} o(1) = 0$.  $H\in \RR^{d\times d}$ is the Hadamard matrix.  $D_b \in \RR^{d\times d}$ is a diagonal matrix whose entries are i.i.d. Rademacher variables.  For a matrix $M \in \RR^{d\times d}$, $M_S$ will denote the restriction of $M$ to its rows indexed by the set $S \subset \{1,...,d\}$.  The variable ${\cal G}$ will always denote a matrix with i.i.d. standard normal Gaussian entries, where the matrix may vary in size.  The variable $\widehat{{\cal G}}$ will always denote a matrix with i.i.d.  copies of a discrete random variable $X$ which roughly models a Gaussian. $C$ will denote various constants that are bounded independent of the dimension.  We will use $m$ to denote the projected dimension of our points, where $m \ll d$, and $d^{\prime}$ the lifted dimension, where $m \leq d^{\prime} \leq d$.  For a vector $x \in S^{d-1}$ we will denote $\tilde{x} = H_S D_b x$.

\section{Main Results }
We now formalize the intuition about how our scheme behaves relative to cross-polytope LSH.
\begin{theorem} \label{fastlsh_theorem}
Suppose $\mathcal{H}$ is the family of hash functions defined in (\ref{fastcphash_eq}) with the choice $m = \mathcal{O}( \ln^5 (d) \ln^4(\ln d))$, and $\rho$ is as defined in (\ref{def_rho}) for this particular family.  Then we have\\
(i-)
\[
\rho = \frac{1}{c^2} \frac{4-c^2 R^2}{4 - R^2} + o(1).
\]
and this hashing scheme runs in time $\mathcal{O} (d  \ln d )$.\\
\\
Moreover, we have the optimal rate of convergence,\\
(ii-)
\[
 \rho = \frac{1}{c^2} \frac{4-c^2 R^2}{4 - R^2} + \mathcal{O} \left(\frac{1}{\ln d^{\prime}} \right).
\]
\end{theorem}
The lower bound given by Theorem 2 in \cite{pracoptlsh} verifies the above rate of convergence is in fact optimal.  We remark that when hashing $n$ points simultaneously, the embedded dimension $m$ picks up a factor of $\ln(n)$.  Assuming that $n$ is polynomial in $d$, the result in Theorem \ref{fastlsh_theorem} still holds simultaneously over all pairs of points.

In addition to creating a fast hashing scheme, one can reduce the amount of randomness involved.  In particular, we show that a slight alteration of the scheme still achieves the optimal $\rho$-value while using only $\mathcal{O}(\ln^9 d)$ bits of randomness.  The idea is to replace the Gaussian matrix by a matrix of i.i.d. discrete random variables.  Some care is required in tuning the size of this matrix so that the correct number of bits is achieved.  As a consequence the number of hash values for this scheme is of order $\mathcal{O}(m)$ (i.e. we lift up to a smaller dimension), which lowers performance in practice, but does not affect the asymptotic sensitivity $\rho$.  We additionally use a JL transform developed by Kane and Nelson \cite{sparsejl} that only uses $\mathcal{O}(\ln (d) \ln(\ln d))$ bits of randomness.  Specifically, the hash function for this scheme is
\[
h_D(x) = \underset{u=\{\pm e_i\}}{\argmin} \left\| \frac{\widehat{{\cal G}}(Zx) }{ \| \widehat{{\cal G}} (Z x) \|_2} - u \right\|_2
\]
where $\widehat{{\cal G}} \in \RR^{d^{\prime} \times m}$ is a matrix with i.i.d. copies of a discrete random variable $X$ which roughly models a Gaussian, and $Z \in \RR^{d \times m}$ is the JL transform constructed in \cite{sparsejl}.  Our analysis allows us to pick the threshold value $d^{\prime} = m$ to minimize the number of random bits.
\begin{theorem} \label{redrand_theorem}
There is a hash family $\mathcal{H}$ with $\mathcal{O}(\ln^9 d)$ bits of randomness that achieves the bound
\[
\rho = \frac{1}{c^2} \frac{4-c^2 R^2}{4 - R^2} + o(1),
\]
and runs in time $\mathcal{O}(d \ln d)$.
\end{theorem}

\subsection{Theorem \ref{fastlsh_theorem} Part (i-) Proof Outline}
First we state an elementary limit result that we will apply to the proofs of both Theorem \ref{fastlsh_theorem} and Theorem \ref{redrand_theorem}.

\begin{lemma} \label{limit_lemma}
Suppose $m_d(a),m_d(b)$ are positive functions, $\lim_{d\to \infty} m_d(a) = a$, $\lim_{d\to \infty} m_d(b) = b$, and that $f(d),g(d)$ are also positive, $\lim_{d\to \infty} f(d) = \lim_{d\to \infty} g(d) = \infty$, $\lim_{d \to \infty} \frac{f(d)}{g(d)} = \infty$.  Then,

\[
\lim_{d\to \infty} \frac{m_d(a) f(d) + g(d)}{m_d(b) f(d) + g(d)} = \frac{a}{b}
\]
\end{lemma}

Proceeding to the proof of Theorem \ref{fastlsh_theorem}, the key observation is that for $x,y \in S^{d-1}$, $\mathcal{G} \tilde{x} = \mathcal{G}_0 \begin{bmatrix} \tilde{x}\\0 \end{bmatrix}$, where $\mathcal{G}_0 \in \RR^{d^{\prime} \times d^{\prime}}$ is a square Gaussian matrix.  Thus,
\[
\PP[h_f(x) = h_f(y)] = \PP\left[ h\left( \begin{bmatrix} \tilde{x}\\0 \end{bmatrix} \right) = h\left( \begin{bmatrix} \tilde{y}\\0 \end{bmatrix} \right) \right],
\]
recalling that $h_f$ is the fast cross-polytope hash function and $h$ is the standard version.  It then follows that, provided the distance between $\tilde{x}$ and $\tilde{y}$ is close to the distance between $x$ and $y$, we can apply proposition \ref{cpcollprob_theorem} to control the above probability.  We start with a lemma for our chosen JL transform that combines a recent improvement on the \emph{restricted isometry property} (RIP) for partial Hadamard matrices \cite{haviv2015restricted} with a reduction from RIP to Johnson-Lindenstrauss transforms in \cite{Rachel}; we defer the proof to the appendix.
\begin{lemma}\label{jl_lemma}
Suppose $\gamma >0$, $x,y\in S^{d-1}$, $\widetilde{x} = H_S D_b x$, $\widetilde{y} = H_S D_b y$ and $H_S \in \RR^{m \times d}$ is such that $m = \mathcal{O}(\gamma \ln^4 (d) \ln^4(\ln d))$.  Then with probability $1 - \mathcal{O}(d^{-\gamma})$, 
\begin{align} 
\left( 1 - \frac{1}{\ln d}\right) &\leq \|\widetilde{x}\|_2^2 \leq \left( 1+  \frac{1}{\ln d} \right), \label{xhat_ineq} \\
\left( 1 - \frac{1}{\ln d }\right ) &\leq \|\widetilde{y}\|_2^2 \leq \left( 1+  \frac{1}{\ln d}\right ), \label{yhat_ineq} \\
\left(1 - \frac{1}{\ln d }\right) \|x-y\|_2^2 &\leq \|\widetilde{x} - \widetilde{y}\|_2^2 \leq \left(1+  \frac{1}{\ln d }\right) \|x-y\|_2^2 \label{xyhat_ineq}
\end{align}

\end{lemma}
We apply the above lemma with the choice $\gamma = \ln d$ to get that
\begin{equation} \label{ang_dist}
\frac{\|x-y\|_2^2}{\left(1 - \frac{1}{\ln d}\right)}  - \frac{5}{\ln d - 1} \leq \left\| \frac{\widetilde{x}}{\|\widetilde{x}\|_2} - \frac{\widetilde{y}}{\|\widetilde{y}\|_2} \right\|_2^2 \leq \frac{\|x-y\|_2^2}{\left(1 + \frac{1}{\ln d }\right)} + \frac{5}{\ln d + 1}.
\end{equation}
with probability $1 - \mathcal{O}(d^{-\ln d})$.  Combining this fact with proposition \ref{cpcollprob_theorem} we get that
\[
\PP[h_f(x) = h_f(y)] = C (d^{\prime})^{\frac{-\tilde{R}^2}{4-\tilde{R}^2}} \ln^{-1}(d^{\prime}),
\]
where $\tilde{R} = \| \tilde{x} - \tilde{y}\|^2$ (by equation (\ref{ang_dist})) goes to $R$ as $d \to \infty$, and $C$ is bounded in the dimension.  We then apply lemma \ref{limit_lemma} to see that
\begin{align*}
\rho &= \frac{\frac{\tilde{R}^2}{4-\tilde{R}^2} \ln(d^{\prime}) +  \ln \ln (d^{\prime}) + C}{\frac{c^2 \tilde{R}^2}{4 - c^2\tilde{R}^2} \ln(d^{\prime}) + \ln \ln(d^{\prime}) + C}\\
&= \frac{1}{c^2} \frac{4-c^2R^2}{4 - R^2} + o(1).
\end{align*}
We defer the proof of Theorem \ref{fastlsh_theorem} part (ii-) to the appendix.
\\

\subsection{Theorem \ref{redrand_theorem} Proof Outline}

We will use the following result (formulated as an analogue to lemma \ref{jl_lemma}) , due to Kane and Nelson, that reduces the amount of randomness required to perform a JL transform.
\begin{proposition} \label{low_rand_prop}
(Theorem 13 and Remark 14 in \cite{sparsejl}) Suppose $\gamma > 0$, $x,y \in S^{d-1}$.  Then, there is a random matrix $Z \in \RR^{d \times m}$ with $m = \mathcal{O}(\gamma \ln^3(d))$ and sampled with $\mathcal{O}(\gamma \ln^2(d))$ random bits such that with probability $1- \mathcal{O}(d^{-\gamma})$,
\begin{align*} 
\left( 1 - \frac{1}{\ln d}\right) &\leq \| Zx \|_2^2 \leq \left( 1+  \frac{1}{\ln d} \right), \\
\left( 1 - \frac{1}{\ln d }\right ) &\leq \|Zy \|_2^2 \leq \left( 1+  \frac{1}{\ln d}\right ),\\
\left(1 - \frac{1}{\ln d }\right) \|x-y\|_2^2 &\leq \|Z(x-y)\|_2^2 \leq \left(1+  \frac{1}{\ln d }\right) \|x-y\|_2^2
\end{align*}
\end{proposition}

Now we want to construct a hash scheme that uses a Gaussian rotation with which to compare our discretized scheme.  Define
\begin{equation} \label{truncatedhashscheme_eq}
h_D^{\prime} (x) = \underset{u=\{\pm e_i\}}{\argmin} \left\| \frac{ {\cal G}^{\prime} Z x}{\|{\cal G}^{\prime} Z x\|_2} - u \right\|_2, 
\end{equation}
where ${\cal G}^{\prime} \in \RR^{m \times m}$ is a standard i.i.d. Gaussian matrix.  The following elementary lemma gives us a suitable replacement for each Gaussian in the matrix ${\cal G}^{\prime}$.

\begin{lemma} \label{discreterv_lemma}
Suppose $g \sim \mathcal{N}(0,1)$.  Then, there is a symmetric, discrete random variable $X$ taking $2^b$ values such that for any $x\in \RR$,
\begin{equation} \label{discretebounds_eq}
\PP[g \leq x] = \PP[X \leq x] + \mathcal{O}(2^{-b})
\end{equation}
\end{lemma}

The discretized scheme can now be constructed by
\begin{equation} \label{lowrandhashscheme_eq}
h_D(x) = \underset{u=\{\pm e_i\}}{\argmin} \left\| \frac{ \widehat{{\cal G}} Z x }{ \| \widehat{{\cal G}} Z  x\|_2}  - u \right\|_2,
\end{equation}
where the entries of $\widehat{{\cal G}} \in \RR^{d^{\prime} \times m}$ are i.i.d. copies of the random variable $X$ in Lemma \ref{discreterv_lemma}.  Note that each discrete random variable has $b$ bits of randomness, so the hashing scheme has minimial randomness when $d^{\prime} = m$, thus there are $m \times m \times b + \mathcal{O}(\gamma \ln^2 (d)) = \mathcal{O}(\gamma^2 \ln^6 (d) b + \gamma \ln^2(d)) $ bits of randomness.  As we will see, we can choose $\gamma$ and $b$ to be a power of $\ln(d)$ while still achieve the optimal asymptotic $\rho$.  For this we have the following lemma.

\begin{lemma} \label{discretecollisionprob_lemma}
Let $x,y\in \RR^d$ be such that $\|x-y\|_2 = R$, $\widetilde{x} = Z x$, and let $h,h^{\prime}$ be as defined in (\ref{lowrandhashscheme_eq}) and (\ref{truncatedhashscheme_eq}) respectively with $m = \mathcal{O}(\ln^4(d))$, $b = \log_2 (d)$ where $\widetilde{R} = \| \widetilde{x} - \widetilde{y}\|_2$.  Then,
\begin{equation} \label{lncollisionprob_eq}
\ln(\PP[h_D(x) = h_D(y)]) = \ln(\PP[h_D^{\prime}(x) = h_D^{\prime}(y)]) + \mathcal{O}_{\widetilde{R}}(1)
\end{equation}

\end{lemma}
We defer the proof of lemma \ref{discretecollisionprob_lemma} to the appendix, but the idea is as follows.  We can first write
\[
\PP[h^{\prime}_D(x) = h^{\prime}_D(y)] = 2d^{\prime} \PP[h^{\prime}_D(x) = h^{\prime}_D(y) = e_1].
\]
Note that the set $\{h^{\prime}_D(x) = h^{\prime}_D(y) = e_1\} = \{ ({\cal G}^{\prime} \widetilde{x})_1 \geq |({\cal G}^{\prime} \widetilde{x})_2|, ({\cal G}^{\prime} \widetilde{y})_1 \geq |({\cal G}^{\prime} \widetilde{y})_2|\},$ which is the Gaussian measure of a convex polytope, so we can write the above probability as the integral over $m$ intervals of the $m$-dimensional Gaussian probability distribution.  We can then use equation (\ref{discretebounds_eq}) to replace the Gaussian pdf with the discrete Gaussian pdf in each coordinate succesively, and (keeping track of parameters), the lemma follows.

We now run the same argument as in Theorem \ref{fastlsh_theorem} by setting $\gamma = \ln d$, so combining lemma \ref{discretecollisionprob_lemma} and proposition \ref{cpcollprob_theorem} applied to $h^{\prime}_D(x)$, we have that
\begin{align*}
\rho &= \frac{\ln(\PP[h_D(x) = h_D(y)])}{\ln(\PP[h_D(cx) = h_D(cy)])}\\
&= \frac{\ln(\PP[h_D^{\prime}(x) = h^{\prime}(y)]) + \mathcal{O}_{\widetilde{R}}(1)}{\ln(\PP[h_D^{\prime}(cx) = h^{\prime}(cy)]) + \mathcal{O}_{\widetilde{R}}(1)}\\
&= \frac{\frac{R_+^2}{4-R_+^2} \ln(d^{\prime}) +  \ln \ln (d^{\prime} ) + C + \mathcal{O}_{\widetilde{R}}(1)}{\frac{c^2 R_-^2}{4 - c^2R_-^2} \ln(d^{\prime}) + \ln \ln(d^{\prime}) + C +\mathcal{O}_{\widetilde{R}}(1)}\\
&=\frac{\frac{R_+^2}{4-R_+^2} \ln(d^{\prime}) +  \ln \ln (d^{\prime}) + C}{\frac{c^2 R_-^2}{4 - c^2R_-^2} \ln(d^{\prime}) + \ln \ln(d^{\prime}) + C}\\
&= \frac{1}{c^2} \frac{4-c^2R^2}{4 - R^2} + o(1)\text{, by lemma \ref{limit_lemma}}.
\end{align*}
Finally, by our choice of $\gamma$ and $b$ in the above lemma, we know that there are $\mathcal{O}(\ln^9(d))$ bits of randomness.

\section{Open Problems} \label{sec_OpenProblems}
Although we achieve a logarithmic number of bits of randomness in Theorem \ref{redrand_theorem}, there is no reason to believe this is optimal among all hash families.  More generally, given a particular rate of convergence to the optimal asymptotic sensitivity we would like to know the minimal number of required bits of randomness.  Note that by the result in \cite{optlblsh}, for each dimension $d$, $c >0$, and $q>0$, there is some distance $R>0$ such that the sensitivity parameter $\rho \geq \frac{1}{c^2} - \mathcal{O}_q\left( \frac{1}{\ln d}\right)$.  In light of this result, we would like to know, for a given rate of convergence, whether it gets close to the lower bound $\frac{1}{c^2}$ for all sequences of distances $R = R(d)$.  Note that this condition holds for cross-polytope lsh with $f(d) = \mathcal{O}\left(\frac{1}{\ln d}\right)$.
\begin{problem} \label{randbits_prob}
Given a rate of convergence $f(d)$ such that $\lim_{d\to \infty} f(d) = 0$, find the minimal number of bits $\mathcal{O}_f(d)$ such that any hash family $\mathcal{H}$ over the sphere $S^{d-1}$ with $\mathcal{O}_f(d)$ bits of randomness satisfies $\rho = \frac{1}{c^2} + f(d)$ for all sequences $R=R(d)$.
\end{problem}
A more practical question is, given a rate of convergence for $\rho$, what is the fastest one could compute a hash family achieving this rate.
\begin{problem}
Given a rate of convergence $f(d)$ as in Problem \ref{randbits_prob},  find the hash family $\mathcal{H}$ over $S^{d-1}$ such that $\rho = \frac{1}{c^2} + f(d)$ for all sequences $R=R(d)$, that also has the fastest hash computations.
\end{problem}
It would be natural to extend our theoretical analysis to the case of hashing a collection of $n$ points simultaneously.  In this setting, the embedding dimension of the JL matrix would inherit an additive factor depending on $\ln(n)$.  Inspired by the construction in \cite{compressedhash} which first sparsifies the data then exploits the restricted isometry property which applies uniformly over all sparse vectors, we can aim for a construction that doesn't depend on the number of data points.

\section{Numerical Experiments} \label{sect_numerexp}
To illustrate our theoretical results in the low dimensional case, we ran Monte Carlo simulations to compare the collision probabilities for regular cross-polytope LSH as well as the fast and discrete versions for various values of the original and lifted dimension.  We refer to \cite{pracoptlsh} for an in depth comparison of run times for cross-polytope LSH and other popular hashing schemes.

The experiments were run with $N = 20000$ trials.  The discretized scheme used 10 bits of randomness for each entry.  The fast, discrete, and regular cross-polytope LSH schemes exhibit similar collision probabilities for small distances, with fast/discrete cross-polytope having marginally higher collision probabilities for larger distances.  It is clear that as the lifted dimension decreases, the fast and discrete versions have higher collision probabilities at further distances, which decreases the sensitivity of those schemes.\\

\begin{figure}[htp]
\centering
  \begin{minipage}[b]{0.3\textwidth}
    \includegraphics[width=1.8in]{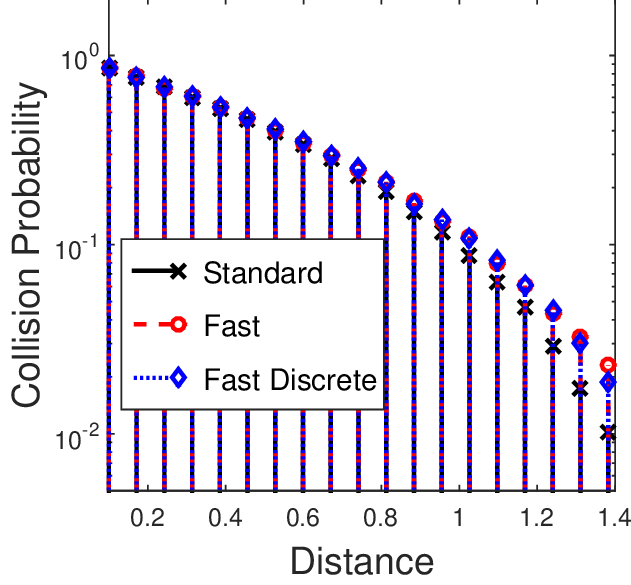}
    \caption{$d=128$,\\ $d^{\prime} = 128$}
  \end{minipage}
  \begin{minipage}[b]{0.3\textwidth}
    \includegraphics[width=1.8in]{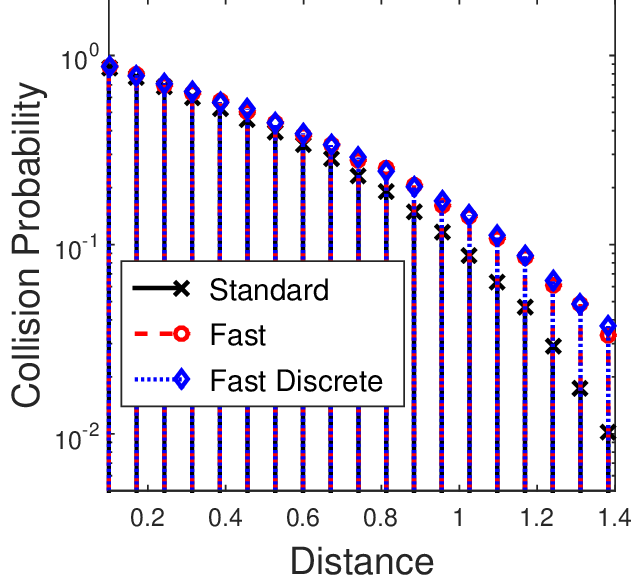}
    \caption{$d=128$,\\ $d^{\prime} = 64$}
  \end{minipage}
  \begin{minipage}[b]{0.3\textwidth}
    \includegraphics[width=1.8in]{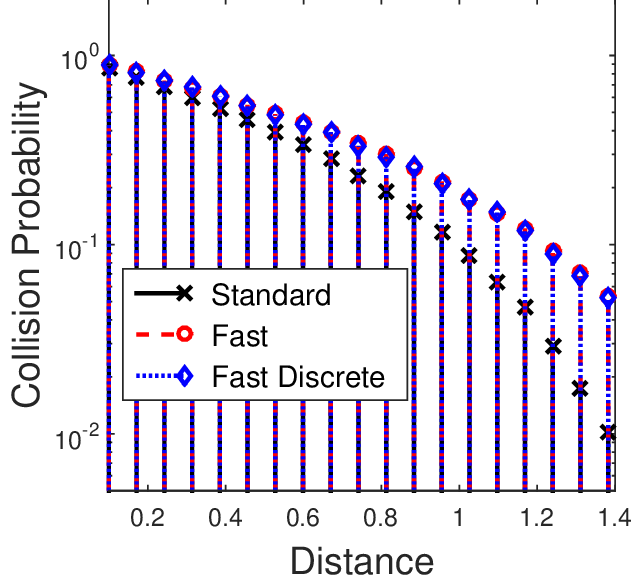}
    \caption{$d=128$,\\ $d^{\prime} = 32$}
  \end{minipage}

\end{figure}

\begin{figure}[htp]
\centering
  \begin{minipage}[b]{0.3\textwidth}
    \includegraphics[width=1.8in]{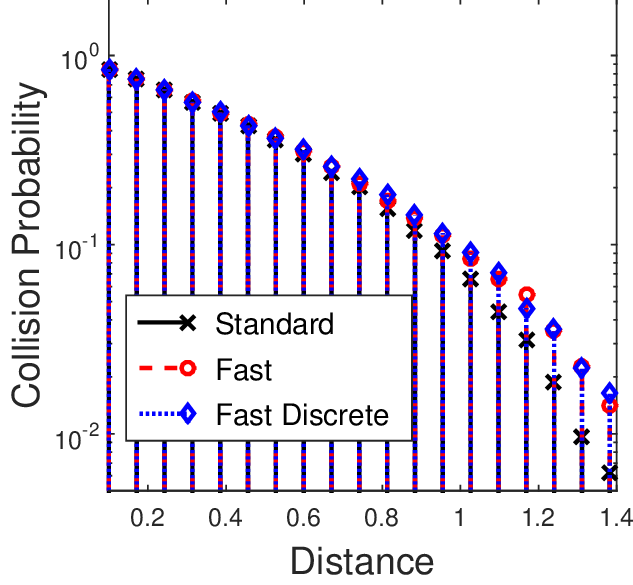}
    \caption{$d=256$,\\ $d^{\prime} = 256$}
  \end{minipage}
  \begin{minipage}[b]{0.3\textwidth}
    \includegraphics[width=1.8in]{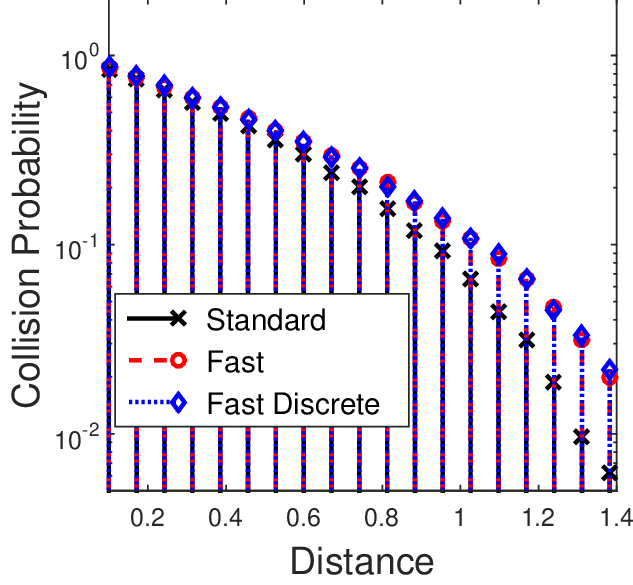}
    \caption{$d=256$,\\ $d^{\prime} = 128$}
  \end{minipage}
  \begin{minipage}[b]{0.3\textwidth}
    \includegraphics[width=1.8in]{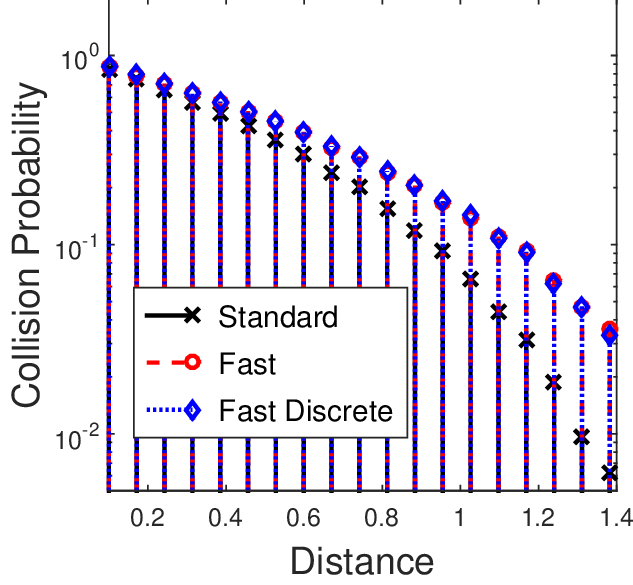}
    \caption{$d=256$,\\ $d^{\prime} = 64$}
  \end{minipage}

\end{figure}
\newpage

The following figures illustrate the rate of convergence to the optimal collision probability as $d\to\infty$, as well as various lines that illustrate the optimal rate of convergence $C\backslash \ln(d)$, where $C$ varies for illustrative purposes.  The experiments were run with varying distances and clearly show the same rate of convergence for the collision probability between the standard and fast cross-polytope schemes.  We note that at low dimensions, the schemes behave even more similarly because the embedded dimension is much closer to the original dimension in this case.

\begin{figure}[htp]
\centering
  \begin{minipage}[b]{0.4\textwidth}
    \includegraphics[width=2.7in]{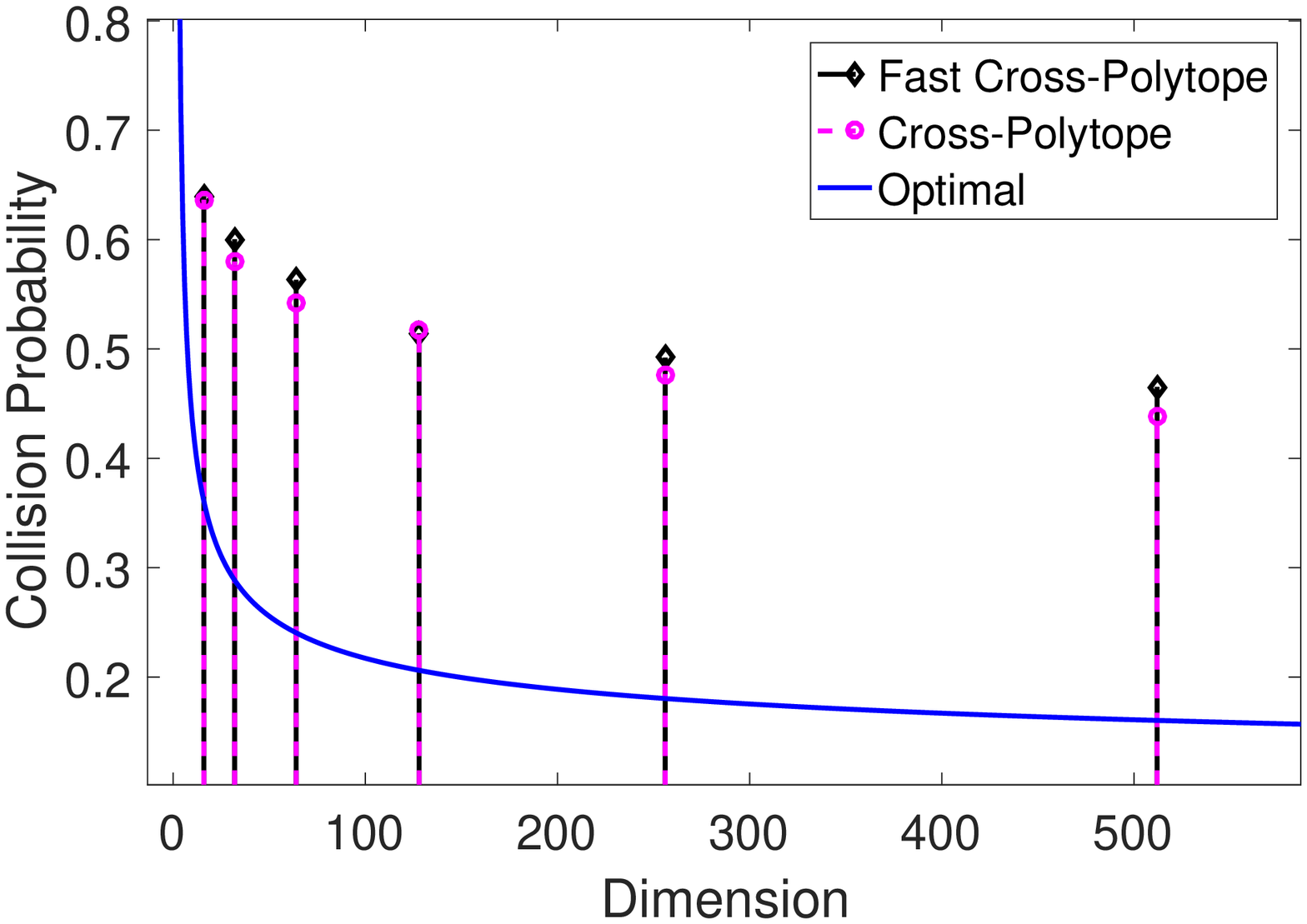}
    \caption{$R = 0.4$}
  \end{minipage}
  \begin{minipage}[b]{0.4\textwidth}
    \includegraphics[width=2.7in]{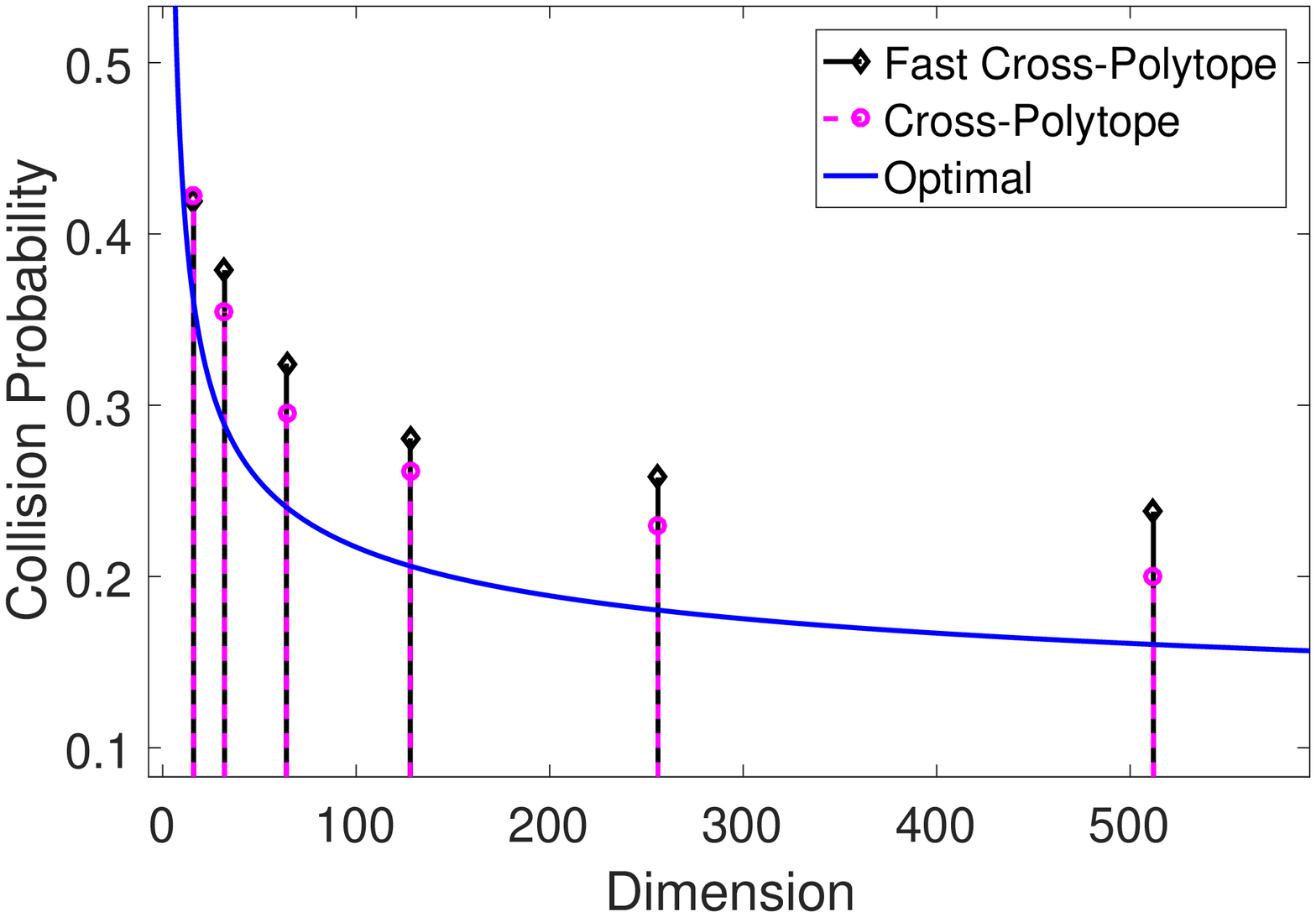}
    \caption{$R = 0.7$}
  \end{minipage}

\end{figure}

\begin{figure}[htp]
\centering
  \begin{minipage}[b]{0.4\textwidth}
    \includegraphics[width=2.7in]{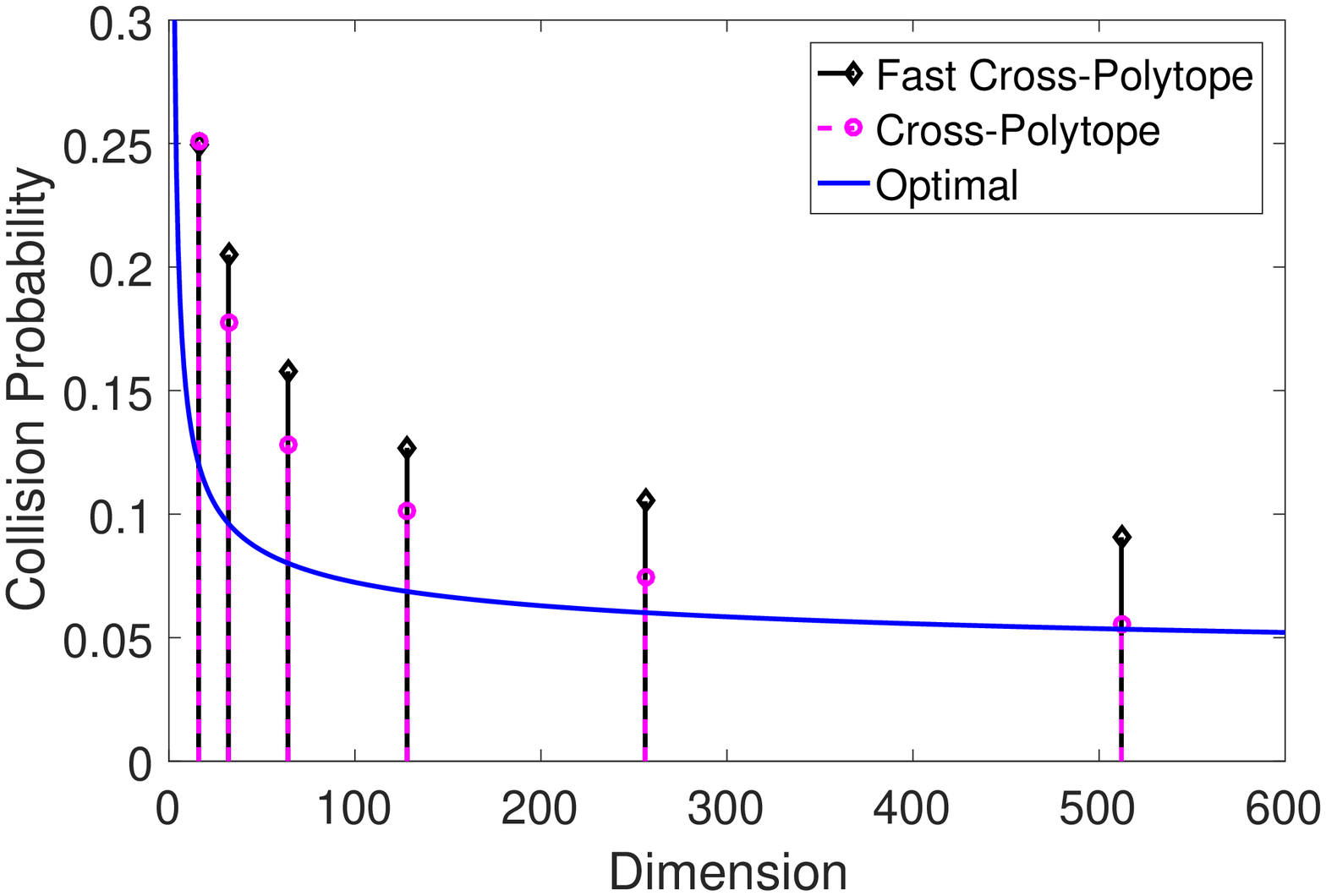}
    \caption{$R = 1$}
  \end{minipage}
  \begin{minipage}[b]{0.4\textwidth}
    \includegraphics[width=2.7in]{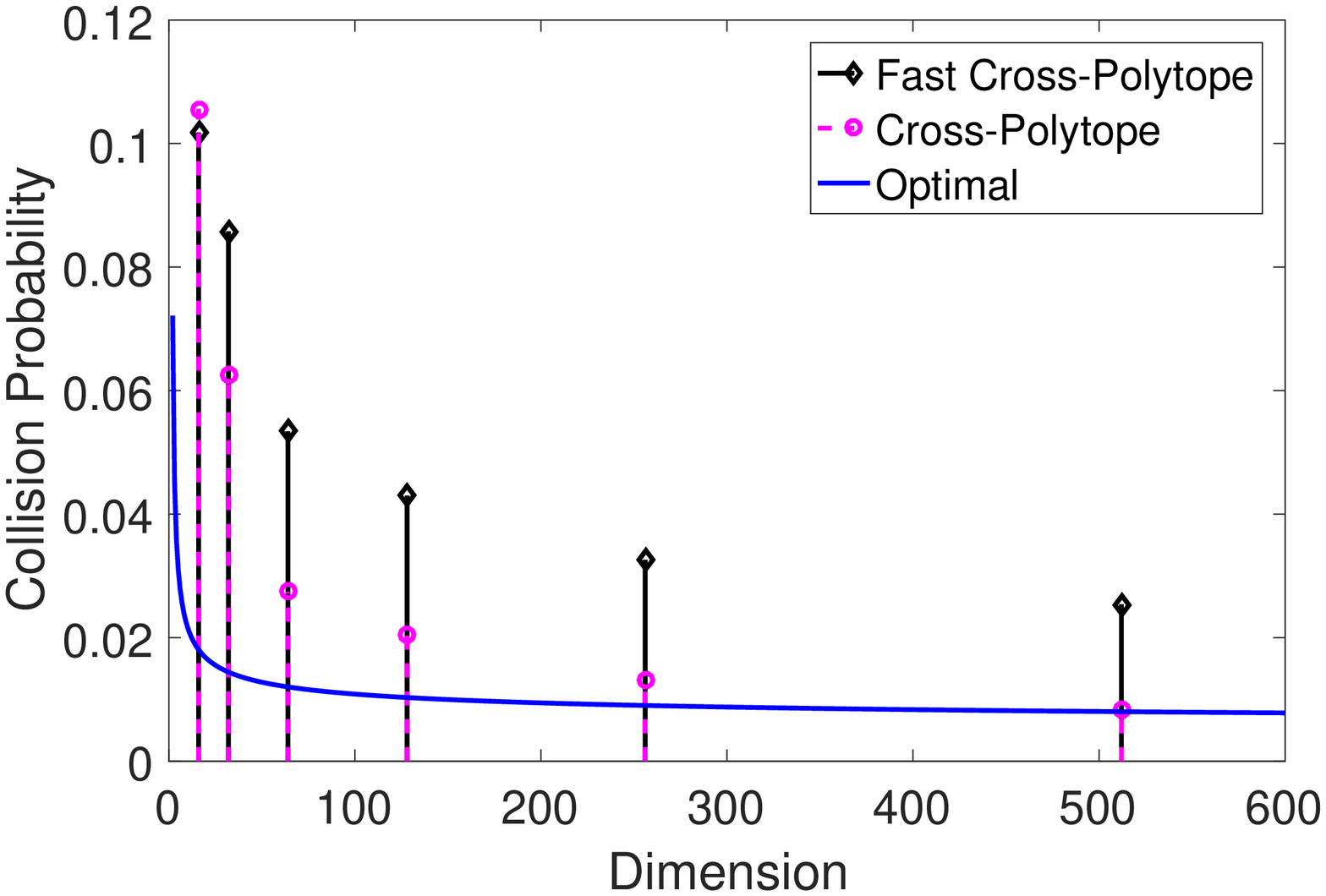}
    \caption{$R = 1.3$}
  \end{minipage}

\end{figure}

\bibliographystyle{alpha}

\bibliography{dimred_bib}

\section{Appendix}

\subsection{Proof of Theorem \ref{fastlsh_theorem} Part (ii-)}

Let $\rho_{R,c}$ be the exponent for standard cross-polytope lsh in dimension $d^{\prime}$, and $\rho^{fast}_{R,c}$ be the exponent for fast cross-polytope lsh lifted to dimension $d^{\prime}$.  Suppose that
\[
\rho_{R,c} - \frac{1}{c^2} \frac{4-c^2R^2}{4-R^2} \leq C(R,c) F(d^{\prime}),
\]
where $F(d^{\prime}) \to 0$ as $d^{\prime} \to \infty$ and $C(r,c)$ is constant in the dimension $d^{\prime}$.\\
Assume that $H_sD_b : \RR^d \to \RR^m$ is a $\delta$-isometry on $x-y$, i.e.
\begin{equation} \label{close_prob}
||x-y||^2_2 \leq R^2 \implies ||\tilde{x} - \tilde{y}||^2_2 \leq (1+\delta) R^2
\end{equation}
\begin{equation} \label{far_prob}
||x-y||^2_2 \geq c^2R^2 \implies ||\tilde{x} - \tilde{y}||^2_2 \geq (1-\delta) c^2 R^2.
\end{equation}
The next observation is that $h_f(x)$ applies the standard cross-polytope lsh scheme on $H_s D_b x$, so conditioned on $H_s D_b x$ being a $\delta$-isometry, we can analyze the fast scheme in terms of the standard scheme as follows:
\[
\rho^{fast}_{R,c} \leq \rho_{R^{\prime},c^{\prime}},
\]
where $R^{\prime} = R \sqrt{1+\delta} $, $c^{\prime} =  \sqrt{\frac{1-\delta}{1+\delta}} c$.  Now, we can say
\begin{align*}
\rho^{fast}_{R,c} - \frac{1}{c^2} \frac{4-c^2R^2}{4-R^2} &\leq [\rho^{fast}_{R,c} - \rho_{R^{\prime}, c^{\prime}}] + \left[\rho_{R^{\prime}, c^{\prime}} - \frac{1}{(c^{\prime})^2} \frac{4 - (c^{\prime})^2 (r^{\prime})^2}{4 - (R^{\prime})^2} \right] + \left[  \frac{1}{(c^{\prime})^2} \frac{4 - (c^{\prime})^2 (R^{\prime})^2}{4 - (R^{\prime})^2} -   \frac{1}{c^2} \frac{4-c^2R^2}{4-R^2} \right]\\
&\leq C(R^{\prime},c^{\prime}) F(d) +  \left[  \frac{1}{(c^{\prime})^2} \frac{4 - (c^{\prime})^2 (R^{\prime})^2}{4 - (R^{\prime})^2} -   \frac{1}{c^2} \frac{4-c^2R^2}{4-R^2} \right].
\end{align*}
The difference in the last equation can be bounded as
\begin{align*}
\frac{1}{(c^{\prime})^2} \frac{4 - (c^{\prime})^2 (R^{\prime})^2}{4 - (R^{\prime})^2} &-   \frac{1}{c^2} \frac{4-c^2R^2}{4-R^2} = \left(\frac{1 + \delta}{c^2(1-\delta)}\right) \frac{4 - (1-\delta)c^2 R^2}{4 - (1-\delta) R^2} - \frac{1}{c^2} \frac{4-c^2R^2}{4-R^2} \\
&\leq \frac{(1+\delta)(4 - (1-\delta)c^2R^2)(4 - R^2) - (4 - c^2 R^2)(1-\delta)(4 - (1-\delta)R^2)}{\frac{c^2}{2} (4 - R^2)^2}\\
&= \delta \mathcal{O}(R,c) + \frac{(1+\delta)(4-c^2R^2)(4-R^2) - (1-\delta)(4-c^2R^2)(4-R^2)}{\frac{c^2}{2} (4 - R^2)^2}\\
&= \delta D(R,c),
\end{align*}
so it follows that $\rho^{fast}_{R,c} - \frac{1}{c^2} \frac{4-c^2R^2}{4-R^2}\leq \delta D(R,c) + C(R^{\prime},c^{\prime}) F(d^{\prime})$ conditioned on the fact that $H_s D_b$ is a $\delta$-isometry on $x-y$.  Note that for $d^{\prime}$ large enough, $C(R^{\prime},c^{\prime})$ is bounded above by a constant independent of the dimension.  We can make the choice $\delta = \frac{1}{\ln(d)}$, so that the isometry condition holds with probability $1- \mathcal{O}(d^{-\ln d})$, so if $\rho$ is the true exponent without conditioning, we get that
\begin{align*}
\rho &\leq \frac{p_1}{p_2 + C \ln\left(1 - d^{-\ln d} \right)}\\
&\leq \frac{p_1}{p_2 - C d^{-\ln d}}\\
&\leq \frac{p_1}{p_2}(1 + C d^{-\ln d}/p_1),
\end{align*}
where $C>0$ is an constant that changes by line but is independent of the dimension.  From this expression it is easy to see that the error term decays at least like $1/\ln d^{\prime}$ (recall that $d^{\prime} \leq d$).\\
Finally, provided $F(d^{\prime})$ decays as fast as than $\frac{1}{\ln(d^{\prime})}$, the result will hold.  This follows from Theorem 1 in \cite{pracoptlsh}.\\

\subsection{Proof of Lemma \ref{limit_lemma}}

We know that for any $\epsilon > 0$ and $d$ large enough, $m_d(b) \geq b - \epsilon$, so that
\begin{align*}
\lim_{d\to \infty} \frac{g(d)}{m_d(b) f(d) + g(d)} &\leq \lim_{d\to \infty} \frac{g(d)}{(b - \epsilon) f(d) + g(d)}\\
&= \lim_{d\to \infty} \frac{1}{(b-\epsilon) \frac{f(d)}{g(d)} + 1} = 0,
\end{align*}
and by positivity the inequality is an equality.  This implies that
\[
\lim_{d\to \infty} \frac{m_d(a) f(d) + g(d)}{m_d(b) f(d) + g(d)} = \lim_{d\to \infty} \frac{m_d(a) f(d)}{m_d(b) f(d) + g(d)}.
\]
The same argument on the reciprocal shows that
\[
\lim_{d\to \infty} \frac{m_d(a) f(d)}{m_d(b) f(d) + g(d)} = \lim_{d\to \infty} \frac{m_d(a) f(d)}{m_d(b) f(d)} = \frac{a}{b}
\]

\subsection{Proof of Lemma \ref{jl_lemma}}
Define the event
\[
E_{v,\delta} := \{ v\in \RR^n : (1-\delta) \|v\|_2 \leq \|\widetilde{v}\|_2 \leq (1+\delta)\|v\|_2 \}.
\]
sCombining Theorem 4.5 of \cite{haviv2015restricted} and Theorem 3.1 of \cite{Rachel}, we know that for any $\eta \in (0,1)$, any $s \geq 40 \ln(12/\eta)$, some $C_0 > 0$, and provided $m = \mathcal{O}(\delta^{-2}\ln^2(1/\delta) s \ln^2(s/\delta) \ln(d))$, 
\[
\PP[E_{x,\delta} \cap E_{y,\delta} \cap E_{x-y,\delta}] \geq (1-\eta)(1 - 2^{-C_0 \ln(d)\ln(s/\delta)})
\]
Setting $\delta = 1/\ln(d)$, $\eta = d^{-\gamma}$, $s = 40 C \ln(12 d)$, we get
\[
\PP[E_{x,\delta} \cap E_{y,\delta} \cap E_{x-y,\delta}] \geq (1 - d^{-\gamma}) (1 - 2^{- C_0 \ln(d) \ln(40\gamma \ln(12d) \ln(d))}),
\]
and the lemma follows.

\subsection{Proof of Lemma \ref{discretecollisionprob_lemma}}

Note that since the entries of $\widehat{{\cal G}} \widetilde{x}$ are symmetric and i.i.d., the probability of hashing to one value is equal for all hash values, so we get
\begin{align}
\PP[h_D (x) = h_D (y)] &= 2d^{\prime} \PP[h_D (x) = h_D (y) = e_1] \nonumber\\
&= 2d^{\prime} \PP[\cap_{j=2}^{d^{\prime}} (\widehat{{\cal G}} \widetilde{x})_1 \geq |(\widehat{{\cal G}} \widetilde{x})_j|, (\widehat{{\cal G}} \widetilde{y})_1 \geq |(\widehat{{\cal G}} \widetilde{y})_j|] \nonumber \\
&= 2d^{\prime} \EE_{(\widehat{{\cal G}} \widetilde{x})_1, (\widehat{{\cal G}} \widetilde{y})_1}( \PP[(\widehat{{\cal G}} \widetilde{x})_1 \geq |(\widehat{{\cal G}} \widetilde{x})_2|, (\widehat{{\cal G}} \widetilde{y})_1 \geq |(\widehat{{\cal G}} \widetilde{y})_2|]^{d^{\prime}-1}) \label{collprob_eq}.
\end{align}
Our goal is to bound the probability $\PP[(\widehat{{\cal G}} \widetilde{x})_1 \geq |(\widehat{{\cal G}} \widetilde{x})_2|, (\widehat{{\cal G}} \widetilde{y})_1 \geq |(\widehat{{\cal G}} \widetilde{y})_2|]$ in terms of the probability $\PP[({\cal G}^{\prime} \widetilde{x})_1 \geq |({\cal G}^{\prime} \widetilde{x})_2|, ({\cal G}^{\prime} \widetilde{y})_1 \geq |({\cal G}^{\prime} \widetilde{y})_2|]$.  Define $E_{{\cal G}^{\prime}} = \{({\cal G}^{\prime} \widetilde{x})_1 \geq |({\cal G}^{\prime} \widetilde{x})_2|, ({\cal G}^{\prime} \widetilde{y})_1 \geq |({\cal G}^{\prime} \widetilde{y})_2|\}$ and similarly for $\widehat{{\cal G}}$.  Since $E_{{\cal G}^{\prime}}$ is a convex polytope, we can write
\begin{align*}
\PP[E_{{\cal G}^{\prime}}] &= \int_{I_1} \int_{I_2(x_1)}...\int_{I_m(x_1,x_2,...,x_{m-1})} \frac{1}{(2\pi)^m} e^{-(x_1^2+...+x_m^2)/2} d x_m ...dx_1,
\end{align*}
where each $I_j(x_1,...,x_j)$ is a (possibly unbounded) interval.  By construction of $X$,
\[
\int_{I_j(x_1,...,x_j)} \frac{1}{2\pi} e^{-x_{j+1}^2/2} dx_{j+1} = \int_{I_j(x_1,...,x_j)} p_X(x_{j+1}) dx_{j+1} +\mathcal{O}( 2^{-b})
\]
where $p_X(x)$ is the pdf of $X$.  This implies that
\begin{align*}
\PP[E_{{\cal G}^{\prime}}] &= \int_{I_1} \int_{I_2(x_1)}...\int_{I_m(x_1,...,x_{m-1})} \frac{1}{(2\pi)^{m-1}} e^{-(x_1^2+...+x_{m-1}^2)/2} p_X(x_m)   dx_{m}...dx_1 + \mathcal{O}(2^{-b})\\
&...= \int_{I_1} \int_{I_2(x_1)}...\int_{I_m(x_1,...,x_{m-1)}} p_X(x_1)...p_X(x_m) dx_m...dx_1 + \mathcal{O}(m 2^{-b})\\
&= \PP[E_{\widehat{{\cal G}}}] +  \mathcal{O}(m 2^{-b}).
\end{align*}
Plugging this into (\ref{collprob_eq}), we get
\begin{align*}
\PP[h_D (x) = h_D (y)] &= 2d^{\prime} \EE_{(\widehat{{\cal G}} \widetilde{x})_1, (\widehat{{\cal G}} \widetilde{y})_1}(\PP[E_{{\cal G}^{\prime}}] + \mathcal{O}(m 2^{-b})))^{d^{\prime} -1}\\
&= 2d^{\prime} \EE_{(\widehat{{\cal G}} \widetilde{x})_1, (\widehat{{\cal G}} \widetilde{y})_1} \left[ \sum_{k=1}^{d^{\prime}-1} \binom{d^{\prime} -1}{k} \PP[E_{{\cal G}^{\prime}}]^k  (\mathcal{O}(m 2^{-b}))^{d^{\prime} - 1 - k}\right ].
\end{align*}
We now make the choice $m = C \ln^4(d)$, $b = \log_2(d) \ln(d)$, so that the above summation becomes
\begin{align*}
\sum_{k=1}^{d^{\prime} -1} \binom{d^{\prime} -1}{k} &\PP[E_{{\cal G}^{\prime}}]^{d^{\prime} -1 - k} (C  \ln^4(d) d^{-\ln(d)})^k \\
&= \sum_{k=1}^{d^{\prime} -1} \binom{d^{\prime} - 1}{k} \PP[E_{{\cal G}^{\prime}}]^{d^{\prime} -1 - k} (C \ln^4(d) d^{-\ln(d)})^{k}
\end{align*}
This first term in the summation is the main term $\PP[E_{{\cal G}^{\prime}}]^{d^{\prime} - 1}$ and the other terms can be bounded using Sterling's approximation as follows,
\begin{align*}
 \binom{d^{\prime}- 1}{k} \PP[E_{{\cal G}^{\prime}}]^{d^{\prime} -1 - k} (C \ln^4(d) d^{-\ln(d)})^{k} &\leq \left( \frac{d^{\prime} e}{k}\right)^k (C \ln^4(d) d^{-\ln(d)})^k.
\end{align*}
For $k \geq 1$ this is certainly bounded by $\mathcal{O}(d^{-\ln(d) + 1})$, and we have
\begin{align*}
\sum_{k=1}^{d^{\prime} -1} \binom{d^{\prime} -1}{k} &\PP[E_{{\cal G}^{\prime}}]^{d^{\prime} -1 - k} (C \ln^4(d) d^{-\ln(d)})^k \\
&= \PP[E_{{\cal G}^{\prime}}]^{d^{\prime} - 1} + \mathcal{O}(d^{-\ln(d) + 2})
\end{align*}
We note that the last asymptotic approximation is very rough but sufficient for our purposes.  This means that
\begin{align}
\PP[h_D (x) = h_D (y)] &= 2d^{\prime} \EE_{(\widehat{{\cal G}} \widetilde{x})_1, (\widehat{{\cal G}} \widetilde{y})_1} (\PP[E_{{\cal G}^{\prime}}]^{d^{\prime} - 1}) + \mathcal{O}(m d^{-\ln(d)+2} ) \label{asycollprob_eq}.
\end{align}
Using the same technique as above where we replace the Gaussian density function with $P_X(x)$, we have
\begin{align*}
\PP[h_D^{\prime} (x) = h_D^{\prime} (y)] &= 2d^{\prime} \EE_{({\cal G}^{\prime} \widetilde{x})_1,({\cal G}^{\prime} \widetilde{y})_1} (\PP[E_{{\cal G}^{\prime}}]^{d^{\prime}- 1})\\
&= 2d^{\prime} \EE_{(\widehat{{\cal G}} \widetilde{x})_1,(\widehat{{\cal G}} \widetilde{y})_2} (\Pr[ E_{{\cal G}^{\prime}}] + \mathcal{O}(m 2^{-b}))^{d^{\prime} - 1}\\
&= 2d^{\prime} \EE_{(\widehat{{\cal G}} \widetilde{x})_1,(\widehat{{\cal G}} \widetilde{y})_2} (\Pr[ E_{{\cal G}^{\prime}}]^{d^{\prime} - 1}) + \mathcal{O}(md^{-\ln(d)+2})
\end{align*}
Finally, plugging this into (\ref{asycollprob_eq}), we get 
\begin{align*}
\PP[h_D (x) = h_D (y)] &= \PP[h_D^{\prime} (x) = h_D^{\prime} (y)] + \mathcal{O}(md^{-\ln(d) + 2})\\
&=\PP[h_D^{\prime} (x) = h_D^{\prime} (y)] + \mathcal{O}(d^{-\ln(d) + 3}). 
\end{align*}
Now, we know that by Theorem \ref{cpcollprob_theorem}, $\ln(\PP[h_D(x) = h_D(y)]) = - \frac{\widetilde{R}^2}{4-\widetilde{R}^2} \ln(d^{\prime}) + \mathcal{O}_{\widetilde{R}}(\ln (\ln d^{\prime}))$, so provided $d$ is large enough that $\ln(d) - 2 > \frac{\widetilde{R}^2}{4 - \widetilde{R}^2}$ , the lemma follows.

\end{document}